\documentclass[aps,pra,reprint,showpacs]{revtex4-1}
\usepackage[dvipdfmx]{graphicx}
\usepackage[dvipdfmx]{color}
\usepackage{amsmath,amssymb}
\usepackage{bm}

\begin{document}


\title{Thermally activated phase slips of one-dimensional
Bose gases in shallow optical lattices}


\author{Masaya Kunimi}
\email{E-mail: masaya.kunimi@yukawa.kyoto-u.ac.jp}
\author{Ippei Danshita}
\affiliation{Yukawa Institute for Theoretical Physics, Kyoto University, Kyoto 606-8502, Japan}



\date{\today}

\begin{abstract}
We study the decay of superflow via thermally activated phase slips in one-dimensional Bose gases in a shallow optical lattice. By using the Kramers formula, we numerically calculate the nucleation rate of a thermally activated phase slip for various values of the filling factor and flow velocity in the absence of a harmonic trapping potential. Within the local density approximation, we derive a formula connecting the phase-slip nucleation rate with the damping rate of a dipole oscillation of the Bose gas in the presence of a harmonic trap. We use the derived formula to directly compare our theory with the recent experiment done by the LENS group [L. Tanzi, et al., Sci. Rep. {\bf 6}, 25965 (2016)]. From the comparison, the observed damping of dipole oscillations in a weakly correlated and small velocity regime is attributed dominantly to thermally activated phase slips rather than quantum phase slips.

\end{abstract}

\maketitle

\section{Introduction}\label{sec:Introduction}

Decay of superflow has been regarded as an important phenomenon in the study of superfluidity since the seminal work done by Landau \cite{Landau1941}. As long as the flow velocity is sufficiently small, a state carrying superflow is metastable due to the existence of an energy barrier, i.e., the superflow persists. However, when the flow velocity exceeds a certain critical value, the energy barrier vanishes and the superflow decays. The critical velocity has been observed in various experiments with ultracold atomic gases \cite{Raman1999,Onofrio2000,Inouye2001,Fallani2004,Sarlo2005,Miller2007,Mun2007,Neely2010,Ramanathan2011,Desbuquois2012,Wright2013,Wright2013_2,Weimer2015}.

The critical-velocity scenario of the decay of superflow has to be modified in low-dimensional systems, especially in one dimension (1D), where the effect of quantum and thermal fluctuations is significant. With such strong fluctuations, a superflow can decay even below the critical velocity due to phase slips (PS), which unwind the phase of the superfluid order parameter to nucleate topological defects, such as solitons or quantum vortices~\cite{Anderson1966}. There are three kinds of PS, any of which emerges depending on the temperature. Near zero temperature, a PS occurs through purely quantum tunneling and is called quantum PS (QPS). When the temperature is rather high but sufficiently low compared to the energy barrier, thermal fluctuations dominate over quantum ones and allow the state to go over the energy barrier. This type of PS is called thermally activated PS (TAPS). In an intermediate temperature, thermal fluctuations are too weak to overcome the energy barrier but so strong that the initial state is mixed with several different states localized at the metastable minimum. This thermal mixing enhances the rate of quantum tunneling and such a PS is called thermally assisted quantum PS (TAQPS). Previous works \cite{Kagan2000,Buchler2001,Khlebnikov2005,Danshita2012,Langer1967,McCumber1970} showed that the nucleation rates of the three kinds of PS exhibit different parameter dependences, such as $\Gamma_{\rm QPS}\sim v^{\alpha}$, $\Gamma_{\rm TAQPS}\sim v T^{\alpha-1}$, and $\Gamma_{\rm TAPS}\sim v e^{-E_{\rm B}/(k_{\rm B}T)}$ for small $v$, where $\alpha$ is a certain constant, $v$ is the flow velocity, $T$ is the temperature, $k_{\rm B}$ is the Boltzmann constant, and $E_{\rm B}$ is the energy barrier.

The superfluid transport of trapped ultracold gases has been often studied through measuring the damping of dipole oscillations (DO) induced by a sudden displacement of the harmonic trapping potential~\cite{Burger2001,Moritz2003,Stoeferle2004,Fertig2005,McKay2008,Haller2010,Gadway2011,Tanzi2013,Boeris2016,Tanzi2016}. In the absence of scatterers, such as impurities and optical lattices, the DO is undamped for many periods~\cite{Burger2001,Moritz2003}. Even in the presence of an optical lattice, a Bose-Einstein condensate in a three-dimensional (3D) (nearly isotropic) trap exhibits undamped DO below the critical velocity~\cite{Burger2001}. In contrast, strong damping has been observed in a one-dimensional (1D) superfluid, which is confined by a strong two-dimensional (2D) optical lattice in the transverse direction, in the presence of the scatterers~\cite{Stoeferle2004,Fertig2005,McKay2008,Haller2010,Gadway2011,Tanzi2013,Boeris2016,Tanzi2016}. It has been suggested that the observed damping may be interpreted as a manifestation of PS~\cite{Danshita2012,Polkovnikov2005}.

Specifically, in the recent experiment by the LENS group, they have measured the velocity dependence of the damping rate in 1D Bose gases in a shallow optical lattice~\cite{Tanzi2016}. They have argued the observation of the crossover from TAQPS to QPS, which is predicted to occur when the velocity increases at sufficiently low temperatures~\cite{Danshita2013}. However, since the experiment has been performed at finite temperatures (specifically, $20\;{\rm nK}\lesssim T \lesssim 50\;{\rm nK}$), it is possible that not only quantum fluctuations but also thermal ones may contribute to some of their experimental results. Hence, in order to clarify whether the observed damping is rooted in quantum or thermal nature, one needs to quantitatively evaluate the rate of TAPS.
          
In this paper, we analyze the effect of TAPS on the damping of DO in 1D Bose gases in a shallow optical lattice. We first consider a system with no harmonic trapping potential to calculate the nucleation rate of TAPS for wide parameter ranges of the filling factor $\nu$ and the velocity $v$ by using the Kramers formula \cite{Langer1969,Hanggi1990}. Within the local density approximation (LDA), we next relate the nucleation rate of PS to the damping rate of DO and compare the latter with the LENS experiment. Our results agree quantitatively with the experiment in a weakly correlated and small velocity region. This indicates that TAPS gives a dominant contribution to the damping in that region over QPS or TAQPS.

The paper is organized as follows: In Sec.~II, we explain the system considered in this paper. In Sec.~III, we describe the method to calculate the nucleation rate. We also derive the relation between the damping rate of DO and the nucleation rate of PS within the LDA. In Sec.~IV, we show our results: the nucleation rates in ring-shaped systems with a shallow optical lattice and the velocity dependence of the damping rates in systems with a trapping potential. In Sec.~V, we summarize our study. In the appendix, we discuss the difference between our results with and without use of the tight-binding approximation.



\section{Setup}\label{subsec:Setup}

In the experiment, the LENS group used ${}^{39}$K atoms and confined them in 1D tubes \cite{Tanzi2016}. The 1D systems were created by using a 2D deep optical lattice in the transverse directions to split a 3D Bose-Einstein condensate. The lattice depth for the transverse confinement is about $20E_{\rm R}$, where $E_{\rm R}\equiv \pi^2\hbar^2/(2m d^2)\simeq h\times 4.53\;$kHz is the recoil energy of the system, $m$ is the atomic mass, and $d=532\;$nm is the lattice spacing. The corresponding transverse trap frequency is given by $\omega_{\perp}= 2\pi\times 40$\;kHz. An optical lattice with period $d$ and depth $s=1$ is added in the longitudinal direction, where $s$ is the lattice depth measured by the recoil energy. It is worth emphasizing that $s=1$ is too shallow for the tight-binding approximation to be quantitative (see the appendix for details). The frequency of the harmonic trap in the longitudinal direction is given by $\omega_{\rm L}=2\pi\times 150$\;Hz.

We treat this system as a quasi-1D system. The 1D external potential is written as $U(x)=(1/2)m\omega_{\rm L}^2x^2+s E_{\rm R}\cos^2(\pi x/d)$. We use $s=1$ in the main part of the paper. The 1D two-body interaction can be written as
\begin{align}
V(x-x')\equiv g_{\rm 1D}\delta(x-x'),\label{eq:one-dimensional_contact_interaction}
\end{align} 
where $\delta(\cdot)$ is the Dirac's $\delta$ function, $g_{\rm 1D}\equiv \hbar^2/(m a_{\rm 1D})$ is a 1D coupling constant, and $a_{\rm 1D}$ is a 1D scattering length. The relation between $a_{\rm 1D}$ and the 3D scattering length $a_{\rm 3D}$ is given by \cite{Olshanii1998,Note1}
\begin{align}
a_{\rm 1D}=\frac{a_{\perp}^2}{2a_{\rm 3D}}\left(1-C\frac{a_{\rm 3D}}{a_{\perp}}\right),\label{eq:relation_a1D_and_a3D}
\end{align}
where $a_{\perp}\equiv \sqrt{\hbar/(m\omega_{\perp})}\simeq 80.6\;$nm is a transverse harmonic oscillator length and and $C=1.03\cdots$ is a constant. Notice that the quasi-1D treatment is well justified under the condition that $\hbar \omega_{\perp} \gg \max(k_{\rm B} T, g_{\rm 1D}n_{\rm 1D})$~\cite{Pitaevskii_and_Stringari}. This condition is, indeed, safely satisfied in 1D Bose gases of the LENS experiment, where $k_{\rm B}T/h \lesssim 1.0 \,{\rm kHz}$ and $g_{\rm 1D}n_{\rm 1D}/h \lesssim 5.4 \,{\rm kHz}$.

The strength of the interaction is characterized by the Lieb-Linger parameter $\gamma_{\rm LL}\equiv 1/(n_{\rm 1D}a_{\rm 1D})$, where $n_{\rm 1D}$ is a 1D particle density of the system. This dimensionless parameter represents the ratio between the healing length and the mean-particle distance. The mean-field approximation employed in this paper is valid for small $\gamma_{\rm LL}$ region \cite{Pitaevskii_and_Stringari, Petrov2000}.

\section{Methods}\label{sec:Methods}

In this section, we describe the methods for calculating the damping rate. Before showing the details, we summarize our methods.

Our calculations presented below are based on a mean-field approximation. In the mean-field approximation, the system is described by a complex order parameter $\Psi(x, t)$, which obeys the Gross-Pitaevskii (GP) equation.

First, we neglect the harmonic potential term $(1/2)m\omega_{\rm L}^2x^2$ and consider a ring-shaped system with an optical-lattice potential. In the ring system, we calculate stable and unstable solutions of the stationary GP equation to evaluate the energy barrier. We also calculate the frequency of the unstable mode and the curvature of the energy landscape around the stationary solutions (Sec.~\ref{subsec:Mean-field Calculation}). 

Second, from these results, we can calculate the nucleation rates of the TAPS $\Gamma$ as a function of  the filling factor $\nu\equiv n_{\rm 1D}d$ and the velocity $v$ by the Kramers formula (Sec.~\ref{subsec:Nucleation_Rate_of_Phase_Slips}).

Finally, we extend the relation between the nucleation rate of PS and the damping rate of DO $G$, which is given by $G\propto \Gamma/v$ \cite{Danshita2013}, by using the LDA in order to include the effects of the inhomogeneity of the harmonic trap (Sec.~\ref{subsec:Local_density_approximation}).

Let us compare our methods to those employed in previous works that have studied TAPS of ultracold Bose gases. Polkovnikov et al. obtained the analytical expression of the TAPS for 1D lattice systems on the basis of the Kramers formula \cite{Polkovnikov2005}. Their expression is derived by using the tight-binding approximation and the quantum-rotor model, which correspond to the high-filling region ($\nu \gg 1$) of the Bose-Hubbard model. Mathey et al. used the truncated Wigner approximation to explain the decay of superflow in a ring trap system \cite{Mathey2014}. They showed that the life time of the superflow depends on the temperature. Although the life time is expected to have exponential dependence about the temperature in the case of TAPS, their calculations could not distinguish the exponential and the power law behavior. Compared to these works, our work has the following two advantages. One is that our methods are applicable to the shallow-lattice regime because our calculations are performed for systems in continuum. Thanks to this advantage, our work can be compared with the LENS experiment without use of any fitting parameters. The other advantage over the truncated Wigner approximation is that the exponential dependence of the TAPS rate on the temperature is explicit in our methods because they are based on the Kramers formula.

\subsection{Mean-field Approximation}\label{subsec:Mean-field Calculation}

\begin{figure}[t]
\centering
\includegraphics[width=7.5cm,viewport=50 54 359 266,clip]{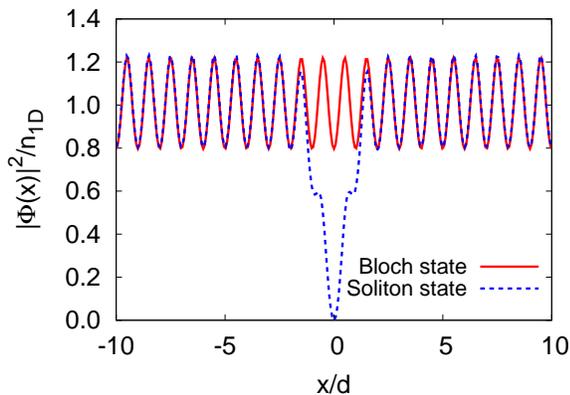}
\caption{(Color online) Density profiles of the Bloch state (red solid line) and the soliton state (blue dotted line) for $\gamma_{\rm LL}=0.19$, $\nu=3$, and $W=0$.}
\label{fig:density_profile}
\end{figure}%

In this subsection, we formulate a ring-shaped system with an optical lattice potential on the basis of the GP mean-field approximation. The ring system satisfies the periodic boundary condition
\begin{align}
\Psi(x+L, t)=\Psi(x, t),\label{eq:periodic_boundary_condition}
\end{align}
where $L\equiv N_{\rm Lat}d$ is the system size and $N_{\rm Lat}$ is the number of lattice sites. We used $N_{\rm Lat}=150$ in this paper. The energy functional of the system is given by
\begin{align}
E&\equiv \int_{-L/2}^{+L/2} dx\left[\frac{\hbar^2}{2m}\left|\frac{\partial\Psi(x, t)}{\partial x}\right|^2+U(x)|\Psi(x, t)|^2\right.\notag \\
&\left.\hspace{14.0em}+\frac{g_{\rm 1D}}{2}|\Psi(x, t)|^4\right].\label{eq:definition_of_energy_functional}
\end{align}
The order parameter $\Psi(x, t)$ is normalized by the particle number $N$:
\begin{align}
N=\int_{-L/2}^{+L/2} dx |\Psi(x, t)|^2.\label{eq:normalization_condition_psi}
\end{align}

The equation of motion of the system is the GP equation:
\begin{align}
i\hbar\frac{\partial}{\partial t}\Psi(x, t)&=\frac{\delta E}{\delta \Psi^{\ast}(x, t)}\notag \\
&=\hspace{-0.25em}\left[-\frac{\hbar^2}{2m}\frac{\partial^2}{\partial x^2}+U(x)+g_{\rm 1D}|\Psi(x, t)|^2\right]\hspace{-0.25em}\Psi(x, t).\label{eq:Gross-Pitaevskii_equation}
\end{align}
The stationary solution of the GP equation is given by $\Psi(x, t)=e^{-i\mu t/\hbar}\Phi(x)$, where $\mu$ is the chemical potential. The periodic boundary condition (\ref{eq:periodic_boundary_condition}) can be rewritten as $\Phi(x+L)=e^{i p L/\hbar}\Phi(x)$, where $p\equiv mv\equiv 2\pi\hbar W/L$ is a crystal momentum and $W\in\mathbb{Z}$ is a winding number. We note that this model has been well investigated in the previous works \cite{Wu2001,Diakonov2002,Menotti2003,Machholm2003,Kramer2003,Wu2003,Seeman2005,Danshita2007,Watanabe2016}.

In the numerical calculations of the GP equation, we use the discrete variable representation method \cite{Baye1986} for the space discretization. We used $2049$ numerical meshes. The ground state solution is obtained by the imaginary time propagation method. Unstable solutions (single soliton solutions, which will be shown below) are obtained by the pseudo-arclength continuation method \cite{Keller1987,Kunimi2015} and the Newton method.

We show typical density profiles of the stable and unstable solutions in Fig.~\ref{fig:density_profile}. The stable solution is a Bloch state, that is, the condition $\Phi(x+d)=e^{i p d/\hbar}\Phi(x)$ is satisfied. On the other hand, the unstable solution is the single soliton solution (blue dotted line in Fig.~\ref{fig:density_profile}), which has a density dip at a maximum of the optical-lattice potential. This solution is $N_{\rm Lat}$-fold degenerate because the center of the density dip may be located on any one of the $N_{\rm Lat}$ maxima of the potential.

\begin{figure}[t]
\centering
\includegraphics[width=7.5cm,viewport=52 52 355 266,clip]{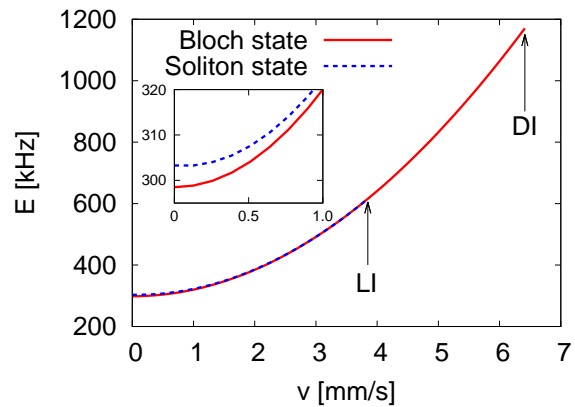}
\caption{(Color online) Velocity dependence of the total energy of the Bloch state (red solid line) and the soliton state (blue dotted line) for $\gamma_{\rm LL}=0.19$ and $\nu=3$. The arrows indicate the points at which the Landau instability (LI) and the dynamics instability (DI) set in. The inset shows a magnified view around a small velocity region.}
\label{fig:energy_diagram}
\end{figure}%

\begin{figure}[t]
\centering
\includegraphics[width=7.5cm,viewport=51 52 359 266,clip]{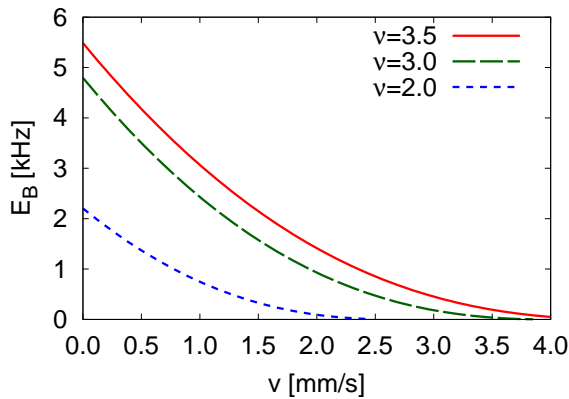}
\caption{(Color online) Velocity dependence of the energy barrier for $\gamma_{\rm LL}=0.19$, $\nu=3.5$ (red solid line), $\nu=3.0$ (green dashed line), and $\nu=2.0$ (blue dotted line).}
\label{fig:energy_barrier}
\end{figure}%

A typical velocity dependence of the total energy is shown in Fig.~\ref{fig:energy_diagram}. The energy barrier vanishes at the point where Landau instability sets in. When the velocity increases further, the dynamical instability sets in~\cite{Wu2003}. Figure~\ref{fig:energy_barrier} shows that typical velocity dependence of the energy barrier. The energy barrier is defined by the energy difference between the soliton state and the Bloch state.  In uniform systems, the energy of the soliton can be analytically obtained \cite{Pitaevskii_and_Stringari}. The energy of the soliton at rest is given by $E_{\rm soliton}=(4/3)\hbar n_{\rm 1D}\sqrt{g_{\rm 1D}n_{\rm 1D}/m}\propto \sqrt{\gamma_{\rm LL}}n_{\rm 1D}^2$. We expect that the same scaling law with respect to the interaction strength and the particle density holds even in the presence of the periodic potential as long as the healing length is sufficiently large compared to the lattice spacing.

\begin{figure}[t]
\centering
\includegraphics[width=7.5cm,viewport=52 52 359 266,clip]{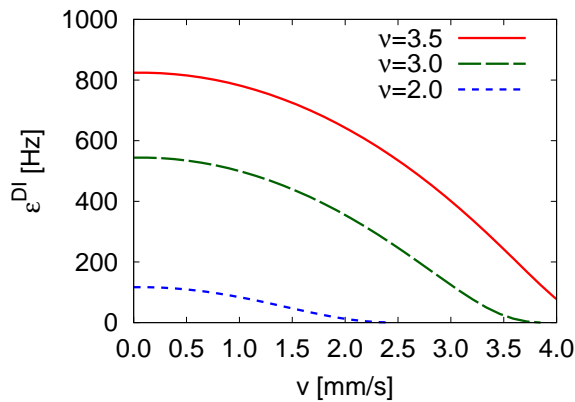}
\caption{(Color online) Frequency of the unstable mode versus the velocity for $\gamma_{\rm LL}=0.19$, $\nu=3.5$ (red solid line), $\nu=3.0$ (green dashed line), and $\nu=2.0$ (blue dotted line).}
\label{fig:growth_rate}
\end{figure}

To obtain excitation spectra, we linearize the GP equation around the stationary solution. This leads to the Bogoliubov equation,
\begin{align}
\!\!\! H_{\rm B}
\begin{bmatrix}
u(x)\\
v(x)
\end{bmatrix}
&=\epsilon
\begin{bmatrix}
u(x)\\
v(x)
\end{bmatrix}
,\label{eq:Bogoliubov_equation}\\
H_{\rm B}&\equiv 
\begin{bmatrix}
\mathcal{K} & g_{\rm 1D}\Phi(x)^2 \\
-g_{\rm 1D}[\Phi^{\ast}(x)]^2 & -\mathcal{K}^{\ast}
\end{bmatrix}
,\label{eq:definition_of_Bogoliubov_matrix}\\
\mathcal{K}&\equiv -\frac{\hbar^2}{2m}\frac{d^2}{d x^2}+U(x)-\mu+2g_{\rm 1D}|\Phi(x)|^2,\label{eq:definition_of_K}
\end{align}
where $\epsilon$ is an excitation energy and $u(x)$ and $v(x)$ are eigenfunctions. A complex $\epsilon$ means that the system is dynamically unstable. For later use, we plot the typical velocity dependence of frequency of the dynamically unstable mode of the soliton state in Fig.~\ref{fig:growth_rate}. We note that the soliton state has the only one unstable mode.

In order to calculate the nucleation rate by the Kramers formula, we must diagonalize the energy matrix that is defined by \cite{Wu2003}
\begin{align}
H_{\rm E}&\equiv \sigma_zH_{\rm B},\label{eq:definition_of_energy_matrix}\\
\sigma_z&\equiv 
\begin{bmatrix}
+1 & 0\\
0 & -1
\end{bmatrix}
.\label{eq:definition_of_sigma3_matrix}
\end{align}
Because the energy matrix is a Hermitian matrix, all of its eigenvalues $\{\lambda_n\}$ are real, in contrast to the Bogoliubov matrix $H_{\rm B}$. The origin of the energy matrix comes from the second-order expansion of the energy functional with respect to the order parameter around the stationary solution. This means that the eigenvalue $\lambda$ can be interpreted as a curvature of the energy landscape around the stationary point.

\subsection{Nucleation Rate of Phase Slips}\label{subsec:Nucleation_Rate_of_Phase_Slips}

\begin{figure}[t]
\centering
\includegraphics[width=7.5cm,clip]{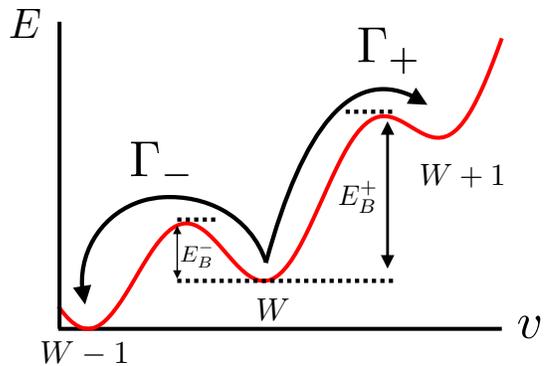}
\caption{(Color online) Schematic picture of TAPS. The red curve represents the energy landscape. $\Gamma_{\pm}$ and $E_{\rm B}^{\pm}$ are the decay rate and the energy barrier of the decay and the acceleration process, respectively.}
\label{fig:energy_landscape}
\end{figure}%

In order to describe effects of thermal fluctuations, we add a noise term and a dissipation term to the GP equation. The dynamics of the system is described by the following Langevin equation:
\begin{align}
(i-\gamma_{\rm dis})\hbar\frac{\partial}{\partial t}\Psi(x, t)&=\frac{\delta E}{\delta\Psi^{\ast}(x, t)}+\mathcal{L}(x, t),\label{eq:Langevin_equation}
\end{align}
where $\gamma_{\rm dis}$ is a dimensionless dissipation constant \cite{Kasamatsu2003}, which is expected to be small in cold atomic gases, and $\mathcal{L}(x, t)$ is a white-Gaussian noise term \cite{McCumber1970}. 

Once the Langevin equation is given, we can derive the equivalent Fokker-Planck equation, which is a partial differential equation for the probability density function. According to the Langer's paper \cite{Langer1969}, we can derive the expression of a decay rate of the metastable state from the Fokker-Planck equation, which is called the Kramers formula \cite{Hanggi1990}. 

\begin{figure}[t]
\centering
\includegraphics[width=7.5cm,viewport=51 52 359 266,clip]{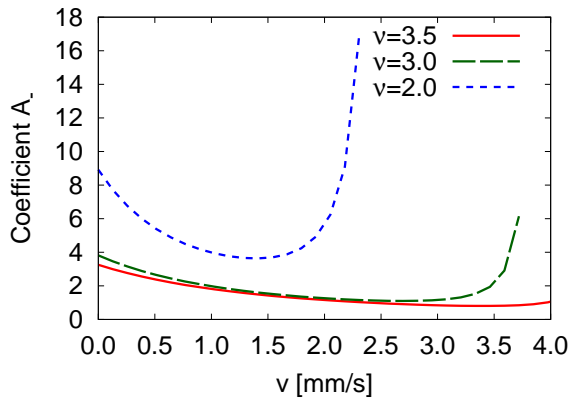}
\caption{(Color online) Velocity dependence of the coefficient $A_-$ for $\gamma_{\rm LL}=0.19$, $\nu=3.5$ (red solid line), $\nu=3.0$ (green dashed line), and $\nu=2.0$ (blue dotted line).}
\label{fig:prefactor}
\end{figure}

Here, we consider two transition processes as shown in Fig.~\ref{fig:energy_landscape}; one changes the winding number from $W$ to $W-1$ (decay process) and the other does from $W$ to $W+1$ (acceleration process), where we assume $W>0$. Nucleation rates for each process $\Gamma_{\pm}$ are given by the Kramers formula: 
\begin{align}
\Gamma_{\pm}&=N_{\rm Lat}\frac{|\kappa^{\pm}|}{2\pi}A_{\pm}e^{-E^{\pm}_{\rm B}/(k_{\rm B}T)},\label{eq:nucleation_rate} \\
A_{\pm}&=\prod_n{}^{'}\sqrt{\frac{\lambda_n^{({\rm m})}}{|\lambda_n^{({\rm u}\pm)}|}},
\end{align}
where the symbol $\pm$ represents the quantity of the decay ($-$) and acceleration ($+$) processes, respectively, $\kappa^{\pm}$ is the frequency of the unstable mode, $\lambda_n^{(\rm m)}$ and $\lambda_n^{({\rm u}\pm)}$ are eigenvalues of the energy matrix, the prime on the product means that we omit the zero modes $(\lambda_n=0)$ in the product, and $E_{\rm B}^{\pm}$ is the energy barrier. The factor $N_{\rm Lat}$ is a consequence of the $N_{\rm Lat}$-fold degeneracy of the soliton solution \cite{Hanggi1990}. The frequency $\kappa^{\pm}$ is obtained by solving the linearized equation of Eq.~(\ref{eq:Langevin_equation}) around the unstable stationary solution $\Psi(x)$:
\begin{align}
(i-\gamma_{\rm dis})\hbar\frac{\partial}{\partial t}\delta\Psi(x, t)=\mathcal{K}\delta\Psi(x, t)+g_{\rm 1D}\Phi(x)^2\delta\Psi^{\ast}(x, t),\label{eq:linearized_equation_for_Langevin_equation}
\end{align}
where we set $\mathcal{L}(x, t)=0$ and $\delta \Psi(x, t)$ is a deviation from the unstable stationary solution. The unstable mode can be written as $\delta\Psi(x, t)=e^{\kappa^{\pm}t}\delta\Phi(x)$. Here, we approximate $\gamma_{\rm dis}\simeq 0$ \cite{Note_dissipation}. Under this approximation, $\kappa^{\pm}$ reduces to ${\rm Im}(\epsilon^{{\rm DI}\pm})/\hbar$, that is the frequency of the unstable mode for the soliton state as shown in Fig.~\ref{fig:growth_rate}. In Fig.~\ref{fig:prefactor}, we show typical profiles of the coefficient $A_{-}$, which is obtained by diagonalizing the energy matrix (\ref{eq:definition_of_energy_matrix}). We note that the Kramers formula is valid for $E_{\rm B}\gg k_{\rm B}T$. We define the nucleation rate of the state labeled by the winding number $W$ as $\Gamma\equiv \Gamma_--\Gamma_+$.

\subsection{Local Density Approximation for Damping Rate\label{subsec:Local_density_approximation}}

\begin{figure}[t]
\centering
\includegraphics[width=7.5cm,clip]{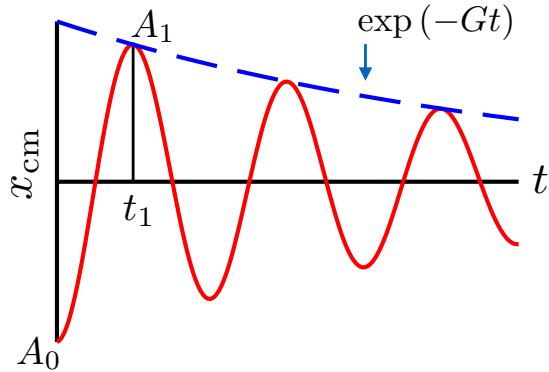}
\caption{(Color online) Schematic picture of the damped oscillation of the center of mass position $x_{\rm cm}$.}
\label{fig:schematic_picture_center_of_mass}
\end{figure}%

In the previous subsection, we described how to calculate the nucleation rates in the ring system with an optical lattice potential and no trapping potential. In this subsection, we derive the relation between the nucleation rate and the damping rate of the DO within the local-density approximation (LDA). 

We first review the relation between the nucleation rate of the PS $\Gamma$ and the damping rate of the DO $G$ \cite{Danshita2013} for the reader's convenience before extending it on the basis of the LDA. To derive the relation between $\Gamma$ and $G$, we calculate the energy loss through the damping of the first half period by the following two ways. The energy loss due to the lost potential energy is given by
\begin{align}
E_{\rm Loss}=N\frac{1}{2}m\omega_{\rm L}^2(A_0^2-A_1^2),\label{eq:energy_loss_potential_energy}
\end{align}
where $A_0$ and $A_1$ are the amplitude of the oscillation at the initial time and the half period $t_1\equiv \pi/\omega_{\rm L}$, respectively (see Fig.~\ref{fig:schematic_picture_center_of_mass}). Assuming the underdamped oscillation ($1-A_1/A_0\ll 1$ or equivalently $G \ll \omega_{\rm L}$), we obtain
\begin{align}
E_{\rm Loss}\simeq N m v_{\rm max}^2G t_1,\label{eq:energy_loss_potential_energy_final_form}
\end{align}
where $v_{\rm max}$ is the maximum velocity of the DO and we used $v_{\rm max}\simeq \omega_{\rm L}A_0$ and $A_1\simeq A_0e^{-G t_1}\simeq A_0(1-G t_1)$.

The other expression of the energy loss is obtained by considering the Joule heat. Let $P$ be a power of the system. The energy loss due to the Joule heat is given by
\begin{align}
E_{\rm Loss}=P\times t_1.\label{eq:energy_loss_Joule_heat}
\end{align}
The power can be written as $P=R I^2$, where $R$ is the resistance of the system and $I$ is the particle current. If we assume that the source of the resistance is phase slips, we can write $R=2\pi\hbar \Gamma/I$ \cite{Langer1967}. The particle current is written as $I\sim n_{\rm 1D} v_{\rm max}$. Therefore, the expression of the energy (\ref{eq:energy_loss_Joule_heat}) can be represented as the nucleation rate:
\begin{align}
E_{\rm Loss}\sim 2\pi\hbar n_{\rm 1D}v_{\rm max}\Gamma t_1.\label{eq:energy_loss_phase_slip_final_form}
\end{align}
Comparing Eqs.~(\ref{eq:energy_loss_potential_energy_final_form}) and (\ref{eq:energy_loss_phase_slip_final_form}), we obtain the relation between $\Gamma$ and $G$:
\begin{align}
G\sim \frac{2\pi\hbar n_{\rm 1D}}{N m}\frac{\Gamma}{v_{\rm max}}.\label{eq:relation_G_and_Gamma}
\end{align}

The above relation includes neither the inhomogeneity of the density profile due to the trap potential nor the temporal change of the velocity. To include these effects, we replace $P$ with the power per unit length $\tilde{P}$ at time $t$:
\begin{align}
&P\to \tilde{P}[n(x, t), v(t; G)]\notag \\
&\quad=2\pi\hbar\tilde{\Gamma}[n(x, t), v(t;G)]n(x, t)v(t;G),\label{eq:LDA_for_power}
\end{align}
where $\tilde{\Gamma}\equiv \Gamma/L$ is the nucleation rate per unit length. We assume that the spatial and temporal dependences of the power and the nucleation rate stem from the local particle density and the velocity. We also assume that the density and velocity are given by
\begin{align}
n(x, t)&\equiv n_{\rm TF}(x-x_{\rm cm}(t)),\label{eq:density_for_LDA}\\
n_{\rm TF}(x)&\equiv \frac{1}{g_{\rm 1D}}\left(\mu-\frac{1}{2}m\omega_{\rm L}^2x^2\right)\theta(\mu-m\omega_{\rm L}^2x^2/2),\label{eq:Thomas-Fermi_density_LDA}\\
x_{\rm cm}(t)&\equiv\int^t_0dt'v(t';G),\label{eq:center_of_mass_LDA}\\
v(t;G)&\equiv e^{-G t}v_{\rm max}\sin(\omega_{\rm L}t),\label{eq:velocity_of_dipole_oscillation_LDA}
\end{align}
where $n_{\rm TF}(x)$ is the 1D-Thomas-Fermi density profile \cite{Dunjko2001}, $\theta(\cdot)$ is the Heaviside's step function, $x_{\rm cm}(t)$ is the position of the center of mass during the damped DO, and $v(t)$ is the velocity of the damped DO. The energy loss can be written as the integration of the power:
\begin{align}
E_{\rm Loss}=\int^{t_1}_0d t\int dx \;\tilde{P}[n(x, t), v(t;G)].\label{eq:energy_loss_LDA}
\end{align}
From Eq.~(\ref{eq:energy_loss_potential_energy}) and the relation $A_1\simeq A_0e^{-G t_1}$, we can express the energy loss as
\begin{align}
E_{\rm Loss}=\frac{N}{2}mv_{\rm max}^2(1-e^{-2G t_1}).\label{eq:energy_loss_LDA_potential}
\end{align}
Equating Eq.~(\ref{eq:energy_loss_LDA}) and Eq.~(\ref{eq:energy_loss_LDA_potential}), we obtain the equation that relates the nucleation rate $\Gamma$ to the damping rate $G$:
\begin{align}
&\frac{N}{2}mv_{\rm max}^2(1-e^{-2G t_1})=\int^{t_1}_0d t\int dx \;\tilde{P}[n_{\rm TF}(x), v(t;G)].\label{eq:equation_for_determine_G}
\end{align}
Once the dependence of $\Gamma$ on $v$ and $\nu$ is given, the damping rate $G$ can by determined by solving Eq.~(\ref{eq:equation_for_determine_G}).

We remark on the region for the spatial integral of Eq.~(\ref{eq:equation_for_determine_G}). The nucleation rates become negative unphysically in large $v$ or small $\nu$ regions (see Subsec. \ref{subsec:Nucleation_Rate_in_a_Bulk}). We remove the negative $\Gamma$ regions in the integral in Eq.~(\ref{eq:equation_for_determine_G}). Since these regions correspond to the outer parts of the gas in the harmonic trap, this procedure corresponds to evaluating the integral around the central part of the gas, which contains much larger population and gives the main contribution to the damping rate.


\section{Results}\label{sec:Results}

\subsection{Nucleation Rate in a Bulk}\label{subsec:Nucleation_Rate_in_a_Bulk}
\begin{figure}[t]
\centering
\includegraphics[width=8.5cm,viewport=37 52 395 246,clip]{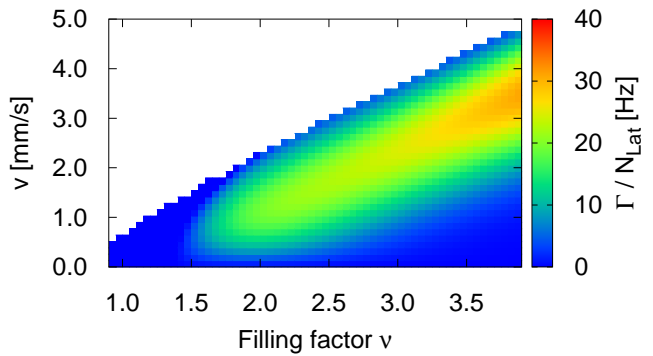}
\caption{(Color online) Nucleation rate per lattice site as a function of the filling factor and the velocity for $\gamma_{\rm LL}=0.19$ and $T=39$ {\rm nK}.}
\label{fig:nucleation_rate_velocity_filling}
\end{figure}

\begin{figure}[t]
\centering
\includegraphics[width=7.5cm,viewport=51 52 353 266,clip]{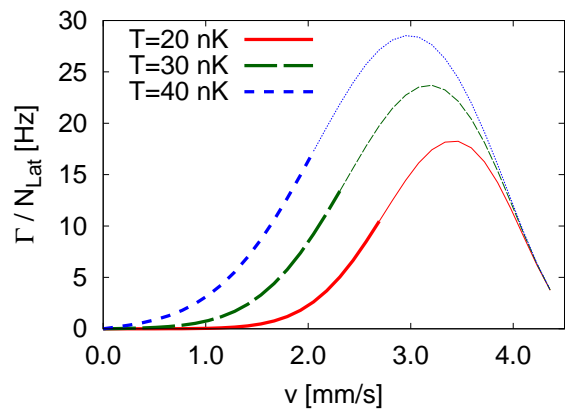}
\caption{(Color online) Nucleation rate per lattice site versus velocity for $\gamma_{\rm LL}=0.19$ and $\nu=3.5$. The red solid, green dashed, and blue dotted lines show the results for $T=20$, $30$, and $40$ {\rm nK}, respectively. The thick lines represent the region $E_{\rm B} > 2k_{\rm B}T$, where the Kramers formula is safely valid.}
\label{fig:nucleation_rate_velocity_temperature}
\end{figure}

We show the results of the nucleation rates of the ring systems. Figure~\ref{fig:nucleation_rate_velocity_filling} shows the nucleation rate per lattice site as a function of the velocity and the filling factor. Moreover, Fig.~\ref{fig:nucleation_rate_velocity_temperature} shows the nucleation rate versus the velocity for the three values of the temperature, where the thin lines indicate the region that $E_{\rm B} \le 2k_{\rm B}T$. As we described in the previous section, the Kramers formula is not reliable in this region. When $v$ is so large or the filling is so low that the system is deep in the invalid regions of the Kramers formula, we find that the nucleation rates become negative, which are not shown in Figs.~\ref{fig:nucleation_rate_velocity_filling} and \ref{fig:nucleation_rate_velocity_temperature}. We emphasize that such an unphysical behavior does not appear when the energy barrier is sufficiently high, i.e., $E_{\rm B} > 2k_{\rm B}T$.

\begin{figure*}[t]
\centering
\includegraphics[width=1.0\linewidth,viewport=49 318 967 736,clip]{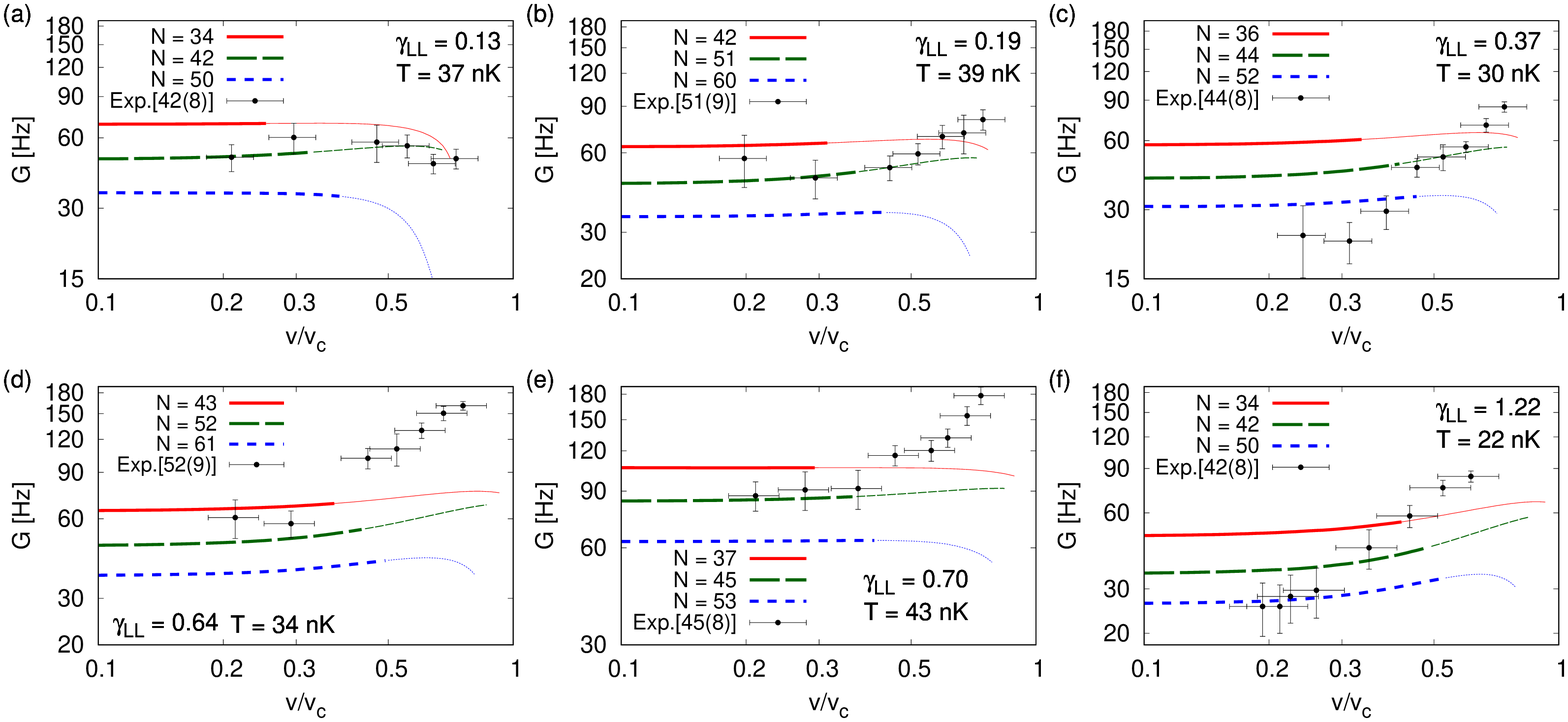}
\caption{(Color online) Damping rates for (a) $(\gamma_{\rm LL},T)=(0.13,37{\rm nK})$, (b) $(\gamma_{\rm LL},T)=(0.19, 39{\rm nK})$, (c) $(\gamma_{\rm LL}, T)=(0.37, 30{\rm nK})$, (d) $(\gamma_{\rm LL},T)=(0.64,34{\rm nK})$, (e) $(\gamma_{\rm LL},T)=(0.70,43{\rm nK})$, and (f) $(\gamma_{\rm LL},T)=(1.22,22{\rm nK})$. The black points represent the experimental data in Ref.~\cite{Tanzi2016}. The thick lines represent the region in which the system at the trap center satisfies $E_{\rm B}> 2k_{\rm B}T$.}
\label{fig:damping_rates_n}
\end{figure*}%

\begin{figure*}[t]
\centering
\includegraphics[width=1.0\linewidth,viewport=49 318 967 736,clip]{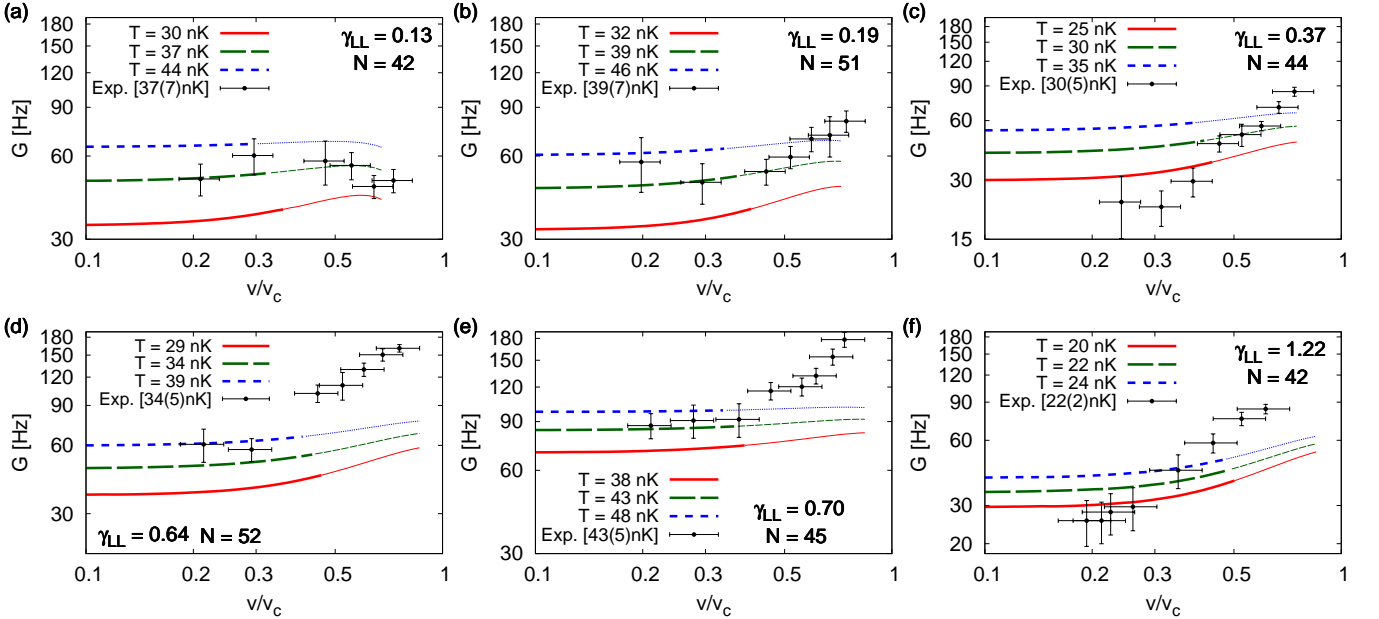}
\caption{(Color online) Damping rates for (a) $(\gamma_{\rm LL},N)=(0.13,42)$, (b) $(\gamma_{\rm LL},N)=(0.19,51)$, (c) $(\gamma_{\rm LL},N)=(0.37,44)$, (d) $(\gamma_{\rm LL},N)=(0.64, 52)$, (e) $(\gamma_{\rm LL},N)=(0.70,45)$, and (f) $(\gamma_{\rm LL},N)=(1.22,42)$. The black points represent the experimental data in Ref.~\cite{Tanzi2016}. The thick lines represent the region in which the system at the trap center satisfies $E_{\rm B}> 2k_{\rm B}T$.}
\label{fig:damping_rates}
\end{figure*}%

\subsection{Damping Rate of a Dipole Oscillation}\label{subsec:Damping_Rate_in_Harmonic_Trap}

In this subsection, we show the results of the damping rates obtained by Eq.~(\ref{eq:equation_for_determine_G}). In order to evaluate the integral in　Eq.~(\ref{eq:equation_for_determine_G}) numerically, we used the spline interpolation for the nucleation rates. Although the experimental results are averaged over all tubes, we calculate the damping rate for the central tube. It is expected that the damping rate
estimated at the central tube gives us the lower bound of the damping
rate for the entire system because the damping rate decreases with increasing the particle number per tube as shown in Fig.~\ref{fig:damping_rates_n}.

In Figs.~\ref{fig:damping_rates_n} and \ref{fig:damping_rates}, we plot the damping rates as functions of the velocity for several values of $(\gamma_{\rm LL}, N)$ and $(\gamma_{\rm LL}, T)$ together with the experimental data from Ref.~\cite{Tanzi2016,Note_scattering_length}, respectively, where $N$ is the number of atoms in the central tube. There the critical velocity for the dynamical instability $v_{\rm c}$, which is evaluated for the ring system with the density corresponding to the trap center, sets the unit of the velocity. Since the uncertainty in the number of atoms and the temperature calibrated in the experiment is rather large, we depict the damping rates for three values of the number of atoms or the temperature in each plot. The middle curve corresponds to the median number of atoms or the median temperature while the upper and lower curves do to the upper and lower bounds of the number of atoms or the temperature. If the experimental data fall within the region between the upper and lower curves, it is fair to state that our theoretical results agree with the experimental ones to the extent that the accuracy of the measurement allows. In the following, we will use the term {\it agreement} in this sense.

Let us briefly review the experimental results of Ref.~\cite{Tanzi2016}. In the experimental data shown in Figs.~\ref{fig:damping_rates_n} and \ref{fig:damping_rates}, one sees that the damping rate is almost independent of the velocity for small $v$ while it grows algebraically with increasing the velocity for relatively large $v$. It was argued in Ref.~\cite{Tanzi2016} that this behavior of the damping rate corresponds to the crossover from TAQPS to pure QPS, which had been predicted in Ref.~\cite{Danshita2013}. We examine this argument by evaluating the contribution of TAPS to the damping rate and comparing it with the experimental data.

Our results in Figs.~\ref{fig:damping_rates_n}(a), \ref{fig:damping_rates_n}(b), \ref{fig:damping_rates_n}(d), \ref{fig:damping_rates_n}(e), \ref{fig:damping_rates}(a), \ref{fig:damping_rates}(b), \ref{fig:damping_rates}(d), and \ref{fig:damping_rates}(e) are in good agreement with the experimental data as long as the velocity is sufficiently small so that the use of the Kramers formula is safely justified. Because our calculations include only the effects of TAPS, this agreement indicates that the dominant contribution to the damping of DO for small $v$ is given by TAPS rather than TAQPS. When the velocity is relatively large, the energy barrier is so small that the Kramers formula is invalid. Hence, we cannot make a direct comparison between our theory and the experiment. However, the energy barrier lower than temperature means that the effect of thermal fluctuations is not negligible and that the damping of DO cannot be purely due to QPS.

In Figs.~\ref{fig:damping_rates_n}(c), \ref{fig:damping_rates_n}(f),\ref{fig:damping_rates}(c) and \ref{fig:damping_rates}(f), we see that the experimental data lie outside the region between the upper and lower curves. In the case of Figs.~\ref{fig:damping_rates_n}(f) and \ref{fig:damping_rates}(f), it seems that $\gamma_{\rm LL}=1.22$ is too large to justify the validity of the GP mean-field theory. In contrast, the disagreements seen in Figs.~\ref{fig:damping_rates_n}(c) and \ref{fig:damping_rates}(c) are unexpected and we do not have a  clear explanation for them.

\begin{figure}[t]
\centering
\includegraphics[width=7.5cm,viewport=52 52 359 261,clip]{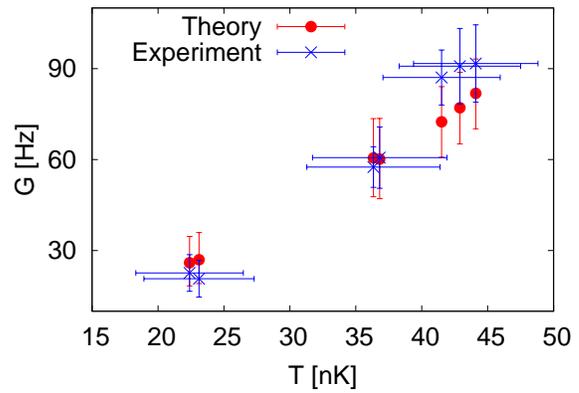}
\caption{(Color online) Damping rate as a function of temperature for $(\gamma_{\rm LL},N)=(0.67, 49)$. The experimental data are taken from the inset of Fig.~5 in Ref.~\cite{Tanzi2016}. The error bars in the theoretical data points are estimated by the value of $G$ at the lower and upper bounds of the experimental temperature.}
\label{fig:damping_rates_temperature_dependence}
\end{figure}%

We next show the temperature dependence of the damping rates in Fig.~\ref{fig:damping_rates_temperature_dependence} together with the experimental data. In the experimental data set, $N$ and $\gamma_{\rm LL}$ are fixed while $v/v_c$ varies within the small-velocity region for different values of temperature. The error bars in the theoretical data points reflect the uncertainty in the experimental temperature. We see that the theoretical results agree with the experimental ones, especially at low temperatures. This result also supports our interpretation that the damping of DO observed in the LENS experiment is attributed dominantly to TAPS when $\gamma_{\rm LL}$ and $v/v_{\rm c}$ are small.

\section{Summary and Discussions}\label{sec:summary_and_discussion}

We investigated the effect of thermal fluctuations on the damping of dipole oscillation of one-dimensional Bose gases in a combined potential of the harmonic trap and optical lattice. Specifically, we aimed to make a direct comparison with the recent experiment done by the LENS group~\cite{Tanzi2016} and to examine their argument that the observed damping is due to quantum fluctuations, i.e., purely quantum phase slips (QPS) or thermally assisted quantum phase slips (TAQPS). To this end, we calculated the nucleation rate of a thermally activated phase slip (TAPS) for a ring-shaped system with a shallow optical lattice by means of the Kramers formula derived on the basis of the Gross-Pitaevskii mean-field approximation. Moreover, we extended the relation between the nucleation rate and the damping rate, which had been derived in Ref.~\cite{Danshita2013}, to a form that includes the spatial inhomogeneity of the density and the temporal dependence of the velocity. Calculating the damping rate through the extended relation, we showed that our theoretical results agree well with the experimental ones when the velocity and the Lieb-Liniger parameter quantifying the strength of quantum fluctuations are small. From this agreement, we argue that the dominant mechanism causing the observed damping is TAPS rather than QPS or TAQPS in the weakly-correlated and small velocity regime.

It is worth emphasizing that from another viewpoint the LENS experiment serves as an observation of superflow decay via TAPS corroborated by quantitative theoretical comparison with no fitting parameter. While the superflow decay via TAPS has been extensively studied both theoretically and experimentally~\cite{Ramanathan2011,McKay2008,Polkovnikov2005,Mathey2014,Kumar2017,Snizhko2016}, none of the previous works has succeeded in making as quantitative a comparison as the present work.

\begin{figure}[t]
\centering
\includegraphics[width=7.5cm,viewport=52 66 355 261,clip]{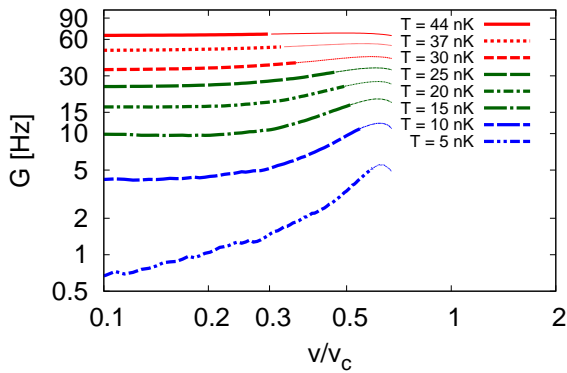}
\caption{(Color online) Damping rate versus velocity for several values of temperature including the ones much lower than achieved in the experiment, where $(\gamma_{\rm LL}, N)=(0.13, 42)$. The thick lines show the region $E_{\rm B}> 2k_{\rm B}T$.}
\label{fig:damping_rate_g013_lower_t}
\end{figure}

Toward observing the damping due to QPS or TAQPS, one needs lower temperature to suppress the contributions of the TAPS. Figure \ref{fig:damping_rate_g013_lower_t} shows the damping rates versus the velocity for several values of temperature including the ones that are lower than achieved in the LENS experiment ($T < 20\,{\rm nK}$). This result suggests that the contribution of the TAPS is well suppressed below $10$\;nK. If one observes larger damping rates than our estimations at sufficiently low temperature, it will be very likely that the dominant contribution to the damping is QPS or TAQPS.

We need further theoretical study in order to exclude or confirm the presence of the QPS in large velocity regions, where the Kramers formula is invalid. If we restrict ourselves to zero temperature, one can use the time-dependent density matrix renormalization method to compute the damping rate even in a system with a shallow optical lattice. If the computed damping rate at zero temperature agrees with the experimental result, it will indicate that QPS is dominant in the observed damping. As an alternative route, let us suggest a possible qualitative way to judge in experiment whether the observed damping is due to QPS or TAPS at a given temperature; namely, measuring the dependence of the damping rate on the central density. According to Ref.~\cite{Danshita2013} [specifically, Fig. 4(a)], the damping rate of the QPS is significantly reduced when the central density is below unity. This happens because the umklapp process, which is the main origin of QPS, is strongly suppressed below the unit filling.  On the other hand, the results in the present paper clearly show that the damping rate of the TAPS increases as the particle number decreases. This qualitative difference between QPS and TAPS may be useful for distinguishing  the QPS from the TAPS.

\begin{acknowledgments}
We thank C. D'Errico for providing us with the experimental data, critically reading the manuscript, and useful comments. M. K. was supported by Grant-in-Aid for JSPS Research Fellow Grant No JP16J07240. I. D. was supported by KAKENHI grants from JSPS, Grants No.~15H05855 and No.~25220711, and by a research grant from CREST, JST.  
\end{acknowledgments}


\appendix
\section{Tight-binding Approximation}\label{app:tight-binding approximation}
The lattice depth in the experiment is $s=1$ \cite{Tanzi2016}, at which the tight-binding approximation does not give correct quantitative results \cite{Pilati2012,Astrakharchik2016}. In this appendix, we compare the nucleation rates in both continuum and lattice systems to show the difference between them. We also check that both results agree at large $s$.

A 1D Bose gas in an optical lattice in the tight-binding regime is approximated by the Bose-Hubbard Hamiltonian,
\begin{align}
\hat{H}&=-J\sum_{j=1}^{N_{\rm Lat}}(\hat{a}_j^{\dagger}\hat{a}_{j+1}+{\rm H.c.})+\frac{U}{2}\sum_{j=1}^{N_{\rm Lat}}\hat{n}_j(\hat{n}_j-1),\label{eq:Bose-Hubbard_Hamiltonian}
\end{align}
where $\hat{a}_j (\hat{a}_j^{\dagger})$ is the annihilation (creation) operator of the boson at the lattice site $j$, $N_{\rm Lat}$ is the total number of the lattice sites, $\hat{n}_j\equiv \hat{a}^{\dagger}_j\hat{a}_j$ is the number operator, $J$ is the hopping parameter, and the $U$ is the onsite interaction. The parameters $J$ and $U$ are numerically calculated by the Wannier function that is obtained by diagonalizing one-particle Schr\"odinger equation with the sinusoidal potential.

We use the mean-field approximation for the lattice system \cite{Smerzi2002}. We replace the annihilation operator $\hat{a}_j$ with the $c$ number $\psi_j(t)$. The GP equation for the lattice system is given by
\begin{align}
i\hbar\frac{d}{d t}\psi_j(t)&=\frac{\partial E}{\partial\psi_j^{\ast}(t)}\notag \\
&=-J\left[\psi_{j+1}(t)+\psi_{j-1}(t)\right]+U|\psi_j(t)|^2\psi_j(t),\label{eq:GP_equation_in_lattice_systems}
\end{align}
where the energy functional for the lattice system is defined by
\begin{align}
E&\equiv\sum_{j=1}^{N_{\rm L}}\left\{-J\left[\psi^{\ast}_j(t)\psi_{j+1}(t)+{\rm c.c.}\right]+\frac{U}{2}|\psi_j(t)|^4\right\}.\label{eq:energy_functional_for_lattice_system}
\end{align}
Because we consider the ring lattice system, $\psi_j(t)$ must satisfy the periodic boundary condition $\psi_{j+N_{\rm Lat}}(t)=\psi_j(t)$. The normalization condition for $\psi_j(t)$ is given by
\begin{align}
\nu&=\frac{1}{N_{\rm Lat}}\sum_{j=1}^{N_{\rm Lat}}|\psi_j(t)|^2.\label{eq:normalization_condition_lattice_system}
\end{align}

The Bloch state solution can be obtained analytically. The solution is given by
\begin{align}
\psi_j(t)&=e^{-i\mu t/\hbar}\sqrt{\nu}e^{i p d j/\hbar},\label{eq:stationary_solution_lattice_system}\\
\mu&=-2J\cos(p d/\hbar)+U\nu,\label{eq:chemical_potential_lattice_system}
\end{align}
where $p\equiv mv\equiv 2\pi \hbar W/(N_{\rm Lat} d)$ is the crystal momentum of the system and $W\in\mathbb{Z}$ is the winding number. The energy of the Bloch states per particle is given by
\begin{align}
\frac{E}{\nu N_{\rm Lat}}&=-2J\cos(p d/\hbar)+\frac{1}{2}U\nu.\label{eq:energy_for_Bloch_state}
\end{align}
To calculate the nucleation rate, we diagonalize the energy matrix. In the lattice system, it is given by
\begin{align}
H_{\rm E}&=
\begin{bmatrix}
\epsilon_{k,p}^{+}-\mu+2U\nu & U\nu \\
U\nu & \epsilon_{k,p}^- -\mu+2U\nu
\end{bmatrix}
,\label{eq:energy_matrix_lattice_system}\\
\epsilon_{k,p}^{\pm}&\equiv -2J\cos(kd\pm p d/\hbar),\label{eq:definition_of_epsilon_pm}
\end{align}
where $k\equiv 2\pi n/(N_{\rm Lat}d) \;(n\in\mathbb{Z})$ is the wave number. The eigenvalues of the energy matrix are given by
\begin{align}
\lambda^{\pm}_k&=\epsilon_k^0\cos(pd/\hbar)+U\nu\notag \\
&\quad \pm\sqrt{4J^2\sin^2(k d)\sin^2(p d/\hbar)+(U\nu)^2},\label{eq:eigenvalue_energy_matrix_lattice_system}\\
\epsilon^0_k&\equiv 4J\sin^2(k d/2),\label{eq:definition_of_single_particle_excitation}
\end{align}
where $\epsilon_k^0$ is the single-particle dispersion of the system.

In contrast to the Bloch states, the soliton states can not be obtained analytically. Therefore, we obtain the soliton solutions by the numerical calculation in the same way as in the case of the continuum system.

\begin{figure}[t]
\centering
\includegraphics[width=7.5cm,viewport=51 52 359 261,clip]{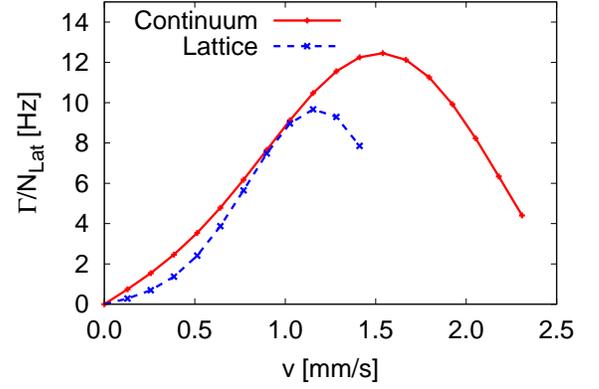}
\caption{(Color online) Nucleation rates per lattice site of the continuum system (red solid line) and the lattice system (blue dotted line) for $s=1$, $\gamma_{\rm LL}=0.19$, and $\nu=2$. This parameter corresponds to $U/J\simeq 0.42$}
\label{fig:nucleation_rate_s=1}
\end{figure}

\begin{figure}[t]
\centering
\includegraphics[width=7.5cm,viewport=51 52 359 261,clip]{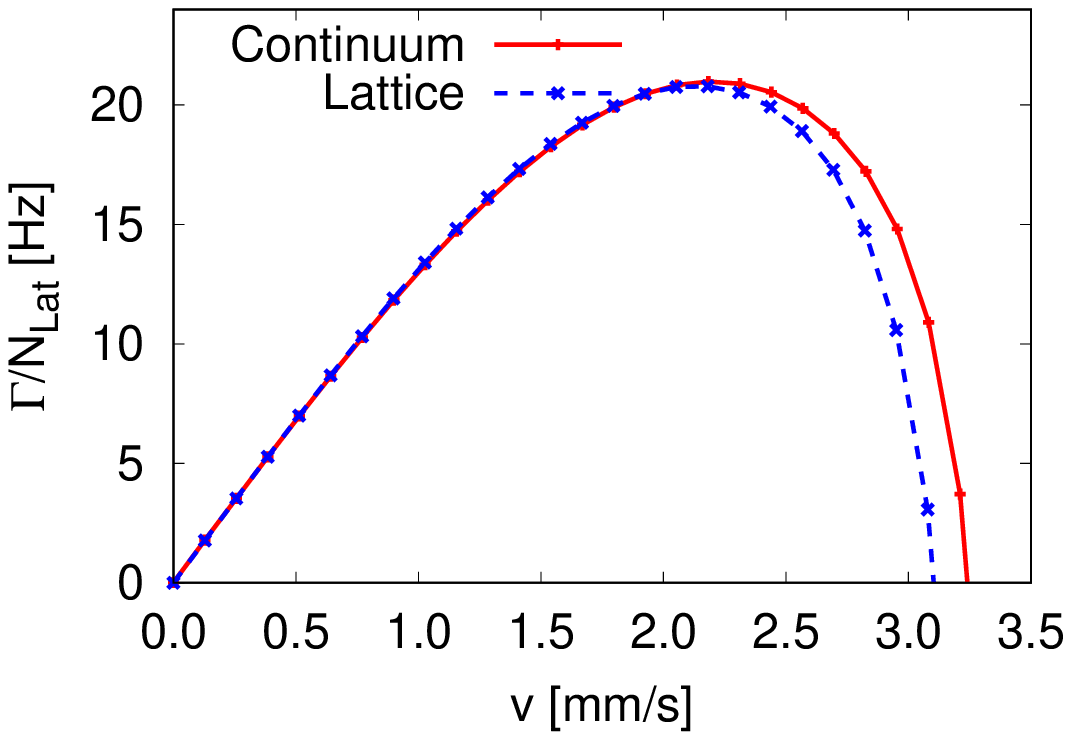}
\caption{(Color online) Nucleation rates per lattice site of the continuum system (red sold line) and the lattice system (blue dotted line) for $s=8$, $\gamma_{\rm LL}=0.19$, and $\nu=2$. This parameter corresponds to $U/J\simeq 3.54$.}
\label{fig:nucleation_rate_s=8}
\end{figure}

Here, we show the results for the nucleation rates of the continuum systems and the lattice systems in Figs.~\ref{fig:nucleation_rate_s=1} and \ref{fig:nucleation_rate_s=8}. We used $N_{\rm Lat}=150$. As shown in Fig.~\ref{fig:nucleation_rate_s=1}, both results clearly deviate. In contrast, as shown in Fig.~\ref{fig:nucleation_rate_s=8} both results are in good agreement when the lattice is sufficiently deep ($s=8$). The deviations in the large-velocity region are due to the contributions of higher band structures.

\newpage

\begin{thebibliography}{99}
\bibitem{Note_scattering_length}
In the LENS experiment, the following 3D-scattering length was used for each $\gamma_{\rm LL}$: $(\gamma_{\rm LL},a_{\rm 3D})=$$(0.13, 25.9(1)a_{\rm B})$, $(0.19, 38.3(2)a_{\rm B})$, $(0.37, 58.0(3)a_{\rm B})$, $(0.64, 93.7(5)a_{\rm B})$, $(0.67, 93.7(5)a_{\rm B})$, $(0.7, 93.7(5)a_{\rm B})$,  and $(1.22, 134.0(9)a_{\rm B})$, where $a_{\rm B}$ is the Bohr radius.
\end{thebibliography}

\end{document}